\pdfoutput=1

\documentclass[%
 reprint,
 floatfix,
 widetext,
 amsmath,amssymb,
 aps,
]{revtex4-2}

\usepackage[toc,page]{appendix}
\usepackage[strict]{changepage}
\usepackage{xcolor}
\usepackage{graphicx}
\usepackage{dcolumn}
\usepackage{bm}

\usepackage[utf8]{inputenc}
\usepackage{physics}
\usepackage{amsfonts}
\usepackage{amsmath}
\usepackage{xcolor}

\begin{document}

\preprint{APS/123-QED}

\title{Exciton Binding Energies in 2D Materials: Insights from Braneworld Physics}

\author{Antoine Honet}
\affiliation{%
Université libre de Bruxelles (ULB), Brussels Laboratory of the Universe (BLU-ULB), Spectroscopy, Quantum chemistry and Atmospheric Remote Sensing (SQUARES), Brussels, Belgium
}%

\author{Michaël Sarrazin}
\affiliation{Universit\'{e}
Marie et Louis Pasteur, CNRS, Institut UTINAM (UMR 6213), \'{E}quipe de Physique Th\'{e}orique, F-25000 Besan\c con, France}

\affiliation{Department of Physics, University of Namur, B-5000 Namur, Belgium}

\date{\today}

\begin{abstract}
In the present work, we introduce a new interpretation of exciton binding energies in two-dimensional (2D) materials using concepts from brane physics. We adapt the Dvali-Gabadadze-Porrati-Shifman mechanism to a (2+1)-dimensional brane in a (3+1)-D spacetime, deriving an effective electromagnetic potential on the brane. Using this potential, we develop a hydrogenic model for exciton binding energies in 2D materials, applying it to s-type excitons and comparing theoretical predictions with experimental results on WS$_{2}$ monolayers. This interdisciplinary approach bridges high-energy and condensed matter physics, offering a new didactic representation of excitons in low-dimensional systems.
 
\begin{description}
\item[Keywords]
DGPS mechanism, quantum electrodynamics, brane, gauge bosons' localization, 2D materials, excitons
\end{description}
\end{abstract}

\maketitle


\author{Antoine Honet and Michaël Sarrazin}

\section{Introduction}
Over the past four decades, there has been a significant interest in braneworld scenarios in theoretical physics. These models describe our visible universe as a 3-dimensional sheet (a 3-brane) embedded in a (3 + 1 + N)-dimensional spacetime ($\mathrm{N} \ge 1$) called the bulk. Such frameworks have provided new approaches to explain phenomena like dark matter and dark energy~\cite{rubakov_we_1983, hughes_supermembranes_1986, lukas_universe_1999, arkani-hamed_manyfold_2000, dvali_quasilocalized_2001, brax_brane_2004, maartens_brane-world_2004, davies_standard_2008, koivisto_dark_2014, bhattacharya_constraining_2017, bhattacharya_cosmological_2018}. 

In this context, the Dvali-Gabadadze-Porrati-Shifman (DGPS) mechanism was introduced in the early 2000s~\cite{dvali_4d_2000, dvali_quasilocalized_2001}. Thanks to the one-loop correction to boson propagators, the DGPS mechanism describes how gauge fields can be localized on a 3-brane embedded in the bulk, assuming fermions' localization on the brane. Then, for instance, a (3+1)-D Newtonian gravity potential ($\propto 1/r$) can be generated on the brane at small distances, while a (4+1)-D potential ($\propto 1/r^2$) is recovered at large distances for a five-dimensional bulk. The same occurs for the electromagnetic field.

However, two-dimensional (2D) materials have attracted a lot of research interest since the experimental isolation of graphene in 2004~\cite{novoselov_electric_2004}. The 2D material family includes, \textit{e.g.}, graphene but also transition metal-dichalcogenide (TMD) monolayers and are studied, among others, for their electronic, magnetic, mechanical or optical properties~\cite{geng_recent_2018}.

In terms of electronic and optical properties, excitons (\textit{i.e.} bound states consisting of an electron and a hole) are of primary importance. In a first approximation, excitons' energies in solids can be described using a hydrogenic model~\cite{kittel_introduction_nodate}, describing the binding energies of excitons as Rydberg series in analogy with hydrogen atom. In this approach, an effective mass has to be used for correcting the energies of the electron-hole system. The Rydberg series for the hydrogen atom is derived using the Coulomb potential in (3+1)-D space-time that is proportional to $1/r$, leading to energies $E_n\propto 1/n^2$ with $n$ the principal quantum number~\cite{cohen-tannoudji_quantum_2020}.

In this paper, in section~\ref{sec:DGPS_3brane}, we recall the framework of the DGPS mechanism in a (4+1)-D bulk involving a 3-brane. Next, in section~\ref{sec:DGPS_2brane}, we adapt the DGPS mechanism to gauge fields' localization on a static 2-brane embedded in a (3+1)-D space-time. In doing so, an effective potential on the brane is derived. Interestingly, this potential coincides with the Rytova-Keldysh potential~\cite{rytova_screened_1967, keldysh_coulomb_1979} derived using 3D electromagnetism and boundary conditions as shown in section~\ref{sec:potential_2_brane}. This potential has been widely applied recently to excitons' physics~\cite{cudazzo_dielectric_2011, pulci_strong_2012, berkelbach_theory_2013, chernikov_exciton_2014, goryca_revealing_2019, hsu_dielectric_2019}. Before to conclude, in section~\ref{sec:excitons_2Dmat}, we show how this model can be successfully used to interpret the properties of excitons in WS$_{2}$ monolayers. 

This work demonstrates the potential for cross-links between high-energy physics and condensed matter theory, illustrating how concepts from physics beyond the Standard Model can be fruitfully applied to solid-state problems. 

Furthermore, our alternative description of the electrostatic potential not only provides a fresh perspective on existing phenomena but also paves the way for generalization to more complex scenarios. These include the consideration of finite-size effects and non-static branes, which may have far-reaching implications for our understanding of emergent phenomena in materials. 

\section{DGPS mechanism for a (4+1)-D space-time}
\label{sec:DGPS_3brane}

\begin{figure}
\includegraphics[width=\columnwidth]{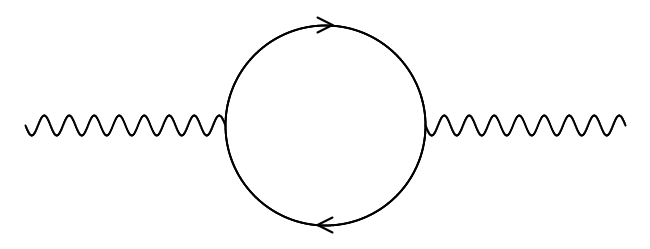}
\caption{Diagram of the one-loop correction to a gauge boson propagator. The wavy lines represent the bare propagator of the boson and the full lines composing the loop represent a fermion propagator.}
\label{fig:diag_1_boucle}
\end{figure}

The basics of the mechanism are the following: if there are fermions localized on a 3-brane embedded in a (4+1)-D space-time, then gauge bosons interacting with the fermions are localized on the brane as well. This is due to the one-loop correction to the boson propagator (see fig.~\ref{fig:diag_1_boucle}). 

More explicitly, considering fermions localized on a static, infinitely thin, 3-brane embedded in a (4+1)-D space-time, the related current is written as~\cite{dvali_quasilocalized_2001}:
\begin{equation}
    J_A(x,y)=J_\mu(x) \delta(y) \delta_A^\mu
    \label{courant}
\end{equation}
where index $A$ represents indices in the (4+1)-D space-time, and $\mu$ represents indices in the (3+1)-hypersurface defined by the evolution of the 3-brane. The coordinate $x$ represents all the coordinates of the hypersurface and $y$ is the fixed coordinate (forced to be $0$ on the brane).

Assuming such a brane current, the gauge boson field $\mathcal{A}_A(x,y)$ can be split in two parts: $\mathcal{A}_y$ and $\mathcal{A}_\mu$. $\mathcal{A}_y$ can be considered as a scalar field decoupled from fermions so that it cannot be observed from the brane. $\mathcal{A}_\mu$ is the effective gauge field interacting with fermions on the brane.

Moreover, an expression for the propagator of the gauge field is found using the one-loop correction~\cite{dvali_quasilocalized_2001}:
\begin{equation}
    D_{\mu \nu}(p)=\frac{\eta_{\mu \nu}}{p^2+2|p|\frac{e^2}{g^2}}
    \label{prop}
\end{equation}
with $g$ the coupling constant of the gauge field in the (4+1)-D space-time and $e$ the effective coupling constant on the brane. $|p|$ is the norm of the quasi-impulsion ($p$) of the boson in the Euclidean space: $|p|=\sqrt{p_0^2+p_1^2+p_2^2+p_3^2}$ and $p^2=|p|^2$. 

This kind of propagator leads to the effective potential on the brane~\cite{dvali_4d_2000}:
\begin{equation}
    V(r)=\frac{-e^2}{2\pi^2} \frac{1}{r} \Bigg[ \bigg(\frac{\pi}{2} -Si(\frac{r}{r_0}) + Ci(\frac{r}{r_0}) sin(\frac{r}{r_0}) \bigg)  \Bigg]
    \label{pot}
\end{equation}
where $r$ is the distance on the brane, $Si$ and $Ci$ the sine and cosine integral functions and $r_0=\frac{g^2}{2e^2}$ depends only on the coupling constants on the bulk and on the brane.\\
The parameter $r_0$ is thus a free parameter of the model, depending on the effective coupling $e$ on the brane. The potential (\ref{pot}) accepts the following limits:

\begin{equation}
    \begin{split}
        &V(r)\simeq -\frac{e^2}{4\pi}\frac{1}{r} \hspace{1.4cm} for \hspace{0.2cm} r<<r_0 \\
        &V(r)\simeq - \frac{g^2}{4\pi^2} \frac{1}{r^2} \hspace{1cm} for \hspace{0.2cm} r>>r_0.
    \end{split}
\end{equation}
In the short distance limit, we recover a potential proportional to $r^{-1}$ as the usual Coulomb potential on a (3+1)-D space-time (i.e. the hypersurface defined by the brane). In the long distance limit, we find a potential proportional to $r^{-2}$ that is typical for a potential in a (4+1)-D space-time (i.e. the bulk). \\
In this (4+1)-D formulation, the DGPS mechanism has been invoked to explain how we could live in a space-time of dimension higher than (3+1) but restricted to a 3-brane so that the laws of physics we experiment are (for distances in the brane small enough) the ones of a (3+1)-D space-time.

\section{DGPS mechanism for a 2-brane in a (3+1)-D space-time}
\label{sec:DGPS_2brane}

In this section, we introduce a model similar to the DGPS approach but canceling a spatial dimension both to the bulk and to the brane in order to describe a 2-brane embedded in a (3+1)-D space-time as model for 2D materials. 

The fermion current associated with localized fermions on the 2-brane is now written as:
\begin{equation}
    J_\mu(x,y)=J_i(x) \delta(y) \delta_\mu^i
    \label{courant2}
\end{equation}
where the $\mu$ index represents the indices of the (3+1)-D bulk and the $i$ index represents the indices of the 2-brane. The $x$ coordinates are the coordinates in the (2+1)-D hypersurface and $y$ is the fixed coordinate of the 2-brane.

As for the initial DGPS mechanism, the vector field associated with the gauge boson $\mathcal{A}_\mu$ can be split into a scalar field $\mathcal{A}_y$ and a vector field $A_i$ on the 2-brane. The scalar field does not interact with fermions so that it cannot be probed from the brane, and we can focus on the vector field $A_i$ only.

The propagator of this vector field is then similar to the one in the original DGPS mechanism, we find:
\begin{equation}
    D_{ij}(p)=\frac{\eta_{ij}}{p^2+2|p|\frac{f^2}{e^2}}
    \label{prop2}
\end{equation}
with $i, j$ the indices on the (2+1)-D hypersurface, $e$ the coupling constant in the bulk and $f$ the effective coupling constant on the 2-brane. $|p|$ is the norm of the quasi-impulsion ($p$) of the boson in Euclidean space: $|p|=\sqrt{p_0^2+p_1^2+p_2^2}$ and $p^2=|p|^2$. 

Up to now, the parallel with the initial DGPS mechanism is complete. When the fermions are located on the brane, the one-loop correction to the boson's propagator induces its localization on the brane as well. The propagator of the vector field (eq.~(\ref{prop2})) is completely analog the one of eq.~(\ref{prop}). 

However, due to dimension loss compared to the initial DGPS mechanism, the resulting potential from the propagator of eq.~(\ref{prop2}) is different from the one found by the (3+1)-D DGPS mechanism.

As for the initial (4+1)-D DGPS mechanism, the potential for the gauge boson is the same as the one deduced from the scalar case of, derived in section~\ref{sec:potential_2_brane}. The potential associated with the localized gauge boson scales as the one derived in section~\ref{sec:potential_2_brane} (eq.~(\ref{final_potential})), and is given by the following expression when electrodynamics' constants are used:
\begin{equation}
    V(r) = \frac{-e^2}{8r_0}  \bigg[ H_0\bigg( \frac{r}{r_0} \bigg) - Y_0\bigg( \frac{r}{r_0} \bigg)   \bigg]
    \label{eq:potential_electrostatic}
\end{equation}
with $r_0=\frac{e^2}{2f^2}$.

The limiting cases are also studied in section~\ref{sec:potential_2_brane} (see eqs.~(\ref{eq:3D_large_lim_case}) and~(\ref{eq:3D_small_lim_case})) and we find:
\begin{equation}
    \begin{split}
        & V(r) \simeq \frac{-e^2}{4\pi} \frac{1}{r} \hspace{2.1cm}  \mathrm{for} \hspace{0.1cm} r>>r_0  \\
        & V(r) \simeq \frac{e^2}{4\pi r_0} ln\bigg( \frac{r}{r_0} \bigg)   \hspace{1cm} \mathrm{for} \hspace{0.1cm} r<<r_0
    \end{split}
    \label{eq:lim_cases_electr}
\end{equation}
where $-1/M_p^2$ in eqs.~(\ref{eq:3D_large_lim_case}) and~(\ref{eq:3D_small_lim_case}) is to be replaced by $e^2$, since the potentials of eqs.~(\ref{eq:potential_electrostatic}) and~(\ref{final_potential}) differ via this substitution. 

The large-distance limit of eq.~(\ref{eq:lim_cases_electr}) scales as the Coulomb potential in a (3+1)-D spacetime. In contrast, the small distance limit exhibits a logarithmic dependence of the potential, which is the typical dependence of a Coulomb potential in a pure (2+1)-D spacetime~\cite{lapidus_classical_1982, mcdonald2019electrodynamics, boito_maxwells_2020}.

Hence, the fact that for small distances, the potential looks like the one of the space-time related to the $p$-brane and, for long distances, the one of the space-time related to the bulk seems to be general, as in the original DGPS mechanism and in our variation of it.

\section{Derivation of the effective potential on the 2-brane from a scalar field}
\label{sec:potential_2_brane}

\subsection{Definition of the action and Green's function formulation}

 In this section, we derive the effective potential using an alternative, yet complementary approach. We consider that the scalar field is dynamically coupled to the brane through the following phenomenological action:
\begin{equation}
\begin{split}
    S = & M_P^2 \int \dd[3]x \dd y \partial_\mu{\phi(x,y)}\partial^\mu{\phi(x,y)} \\
    &+ m \int \dd[3]x \dd y \delta(y) \partial_i{\phi(x,0)}\partial^i{\phi(x,0)}
\end{split}
\end{equation}
with $M_P$ the (3+1)-D coupling constant and $m$ the coupling constant on the 2-brane. Using the term proportional to $m$, this approach phenomenologically considers the interaction between the scalar field and the brane, without explicitly referring to the previously considered one-loop corrections.

The retarded Green's function equation is then:
\begin{equation}
    \Bigg(  M_P^2 \partial_\mu \partial^\mu + m \delta(y) \partial_i\partial^i   \Bigg) G_R(x,y;0,0)=\delta^3(x)\delta(y),
    \label{green}
\end{equation}
that can be written in Fourier space as:
\begin{equation}
    \Bigg(  M_P^2 (p^2-\partial_y^2) + m p^2 \delta(y)  \Bigg) \Tilde{G}_R(p,y)=\delta(y)
    \label{eq:green_Fourier}
\end{equation}
with $\abs{p} = \sqrt{p^2} = \sqrt{p_0^2+p_1^2+p_2^2}$ the Euclidean norm of the momentum on the 2-brane.

The solution for the Green's function is~\cite{dvali_4d_2000, honet_master_2019}:
\begin{equation}
    \Tilde{G}_R(p,y)= \frac{1}{mp^2+2M_P^2\abs{p}} e^{-p\abs{y}}.
    \label{Green_FT}
\end{equation}

Detailed derivation of this expression is given in the Appendix~\ref{sec:app_green}.

\subsection{Derivation of the potential}

From the Green's function, we can express the potential as:
\begin{equation}
    V(r)=\int G_R(x,y=0;0,0)dt
    \label{potential_GF}
\end{equation}
with $t=x_0$ and $r=\sqrt{x_1^2+x_2^2}$ the distance on the brane.

Taking the inverse Fourier transform of eq.~(\ref{Green_FT}), eq.~(\ref{potential_GF}) rewrites as:
\begin{equation}
    V(r)= \frac{1}{2M_P^2} \int \dd{t} \frac{\dd[3]p}{(2\pi)^3} \frac{e^{ipx}}{r_0p^2 + \abs{p}} 
\end{equation}
with $r_0 =  \frac{m}{2 M_P^2}$.

Writing the 2D position and impulsion vectors $\Vec{x}=(x_1, x_2)$ and $\Vec{p}=(p_1, p_2)$, the (2+1)-D Euclidean scalar product writes $px=p_0t+ \Vec{x} \vdot \Vec{p}$. We also define the 2D norm of $\Vec{p}$ as $\norm{p}^2 = p_1^2 + p_2^2$, leading to $p^2=p_0^2 + \norm{p}^2$. 

First performing the integral over $t$ than over $p_0$ simply leads to a factor $(2\pi)$, with the potential reducing to:
\begin{equation}
    V(r)= \frac{1}{(2\pi)^2 2 M_P^2} \int \dd{t} \dd[2]p \frac{e^{i\Vec{x} \vdot \Vec{p}}}{r_0 \norm{\Vec{p}}^2 + \norm{\Vec{p}}} .
\end{equation}

Going into polar coordinates, we get:
\begin{equation}
    V(r)= \frac{1}{(2\pi)^2 2 M_P^2} \int_0^{+\infty} \int_0^{2\pi}  \frac{e^{i r q cos(\theta)}}{r_0 q^2 + q} q \dd q \dd \theta 
\end{equation}
where $q=\norm{\Vec{p}}$ is the norm of the 2D impulsion on the brane and $\theta$ the angle between the 2D vectors $\Vec{x}$ and $\Vec{p}$.

Using the Bessel function of the first kind of order zero ($J_0(x)=\frac{1}{2\pi} \int_0^{2\pi} e^{ix cos(\theta)} \dd \theta$), the potential writes:
\begin{equation}
    V(r) = \frac{1}{4\pi r_0 M_P^2} \int_0^\infty \frac{J_0 (qr)}{q+\frac{1}{r_0}} \dd q.
    \label{tabulated_integr}
\end{equation}

The integral of eq.~(\ref{tabulated_integr}) is tabulated in Ref.~\cite{gradshteyn_table_2015} and using this tabulated integral, the effective potential on the brane reduces to:
\begin{equation}
    V(r)  = \frac{1}{8r_0 M_P^2} \bigg[  H_0\bigg(\frac{r}{r_0} \bigg) - Y_0\bigg(  \frac{r}{r_0}\bigg) \bigg]
    \label{final_potential}
\end{equation}
where $H_0$ is the zeroth order Struve function and $Y_0$ is the zeroth order Neumann function.

This form of potential is known as the Rytova-Keldysh potential~\cite{rytova_screened_1967, keldysh_coulomb_1979} which was first derived by classical electromagnetic considerations. This potential has recently been applied to exciton binding energy models, for example, through simplified analytical solutions~\cite{galiautdinov_anisotropic_2019, henriques_analytical_2020, nguyen-truong_exciton_2022} or numerical methods~\cite{pulci_strong_2012, berkelbach_theory_2013, chernikov_exciton_2014, goryca_revealing_2019, hsu_dielectric_2019, vaquero_excitons_2020, Henriques:21, kirichenko_influence_2021}, among which the variational approach~\cite{pulci_strong_2012, berkelbach_theory_2013, hsu_dielectric_2019, Henriques:21}.

\subsection{Limiting case: $r>>r_0$}

In the long-range regime, the difference between Struve and Neumann functions accept an asymptotic development~\cite{abramowitz_1970}:
\begin{equation}
    H_0 (x) - Y_0 (x) = \frac{2}{\pi} \bigg[ \frac{1}{x} + O(x^{-3}) \bigg].
\end{equation}

The potential $V(r)$ than have the following approximated expression in the long-range limit:
\begin{equation}
    V(r) \simeq \frac{1}{4\pi M_P^2 } \frac{1}{r}
    \label{eq:3D_large_lim_case}
\end{equation}
which scales as $1/r$, that is the scaling of Coulomb potential in (3+1)-D space-time.

\subsection{Limiting case: $r<<r_0$}

We now consider the short-range limit for the expression of eq.~(\ref{final_potential}). The Struve function is defined as~\cite{abramowitz_1970}:
\begin{equation}
    H_0(x) = \frac{x}{2} \sum_{k=0}^{+\infty} \frac{(-1)^k (\frac{z}{2})^{2k}}{[\Gamma(k+\frac{3}{2})] ^2}
    \label{struve_expression}
\end{equation}
while the Neumann function follows the following asymptotic limit~\cite{abramowitz_1970}:
\begin{equation}
    Y_0(x) \sim \frac{2}{\pi} \bigg[ ln(\frac{x}{2}) + \gamma \bigg]
    \label{neumann_small_arg}
\end{equation}
where $\gamma$ is the Euler-Masceroni constant.

While the Struve function behaves as positive powers, the Neumann function diverges as a logarithm for small arguments. The whole term in eq.~(\ref{final_potential}) than follows the logarithmic behavior in the short-range limit and scales as:
\begin{equation}
    V(r) \sim \frac{-1}{ 2\pi m} ln\bigg(\frac{ r}{r_0}\bigg) .
    \label{eq:3D_small_lim_case}
\end{equation}

\section{Application to excitons of 2D materials}
\label{sec:excitons_2Dmat}

When achieving the analogy between a 2-brane and a 2D material, the (2+1)-D DGPS mechanism can be used to invoke localization of photons in the material as a consequence of fermions' localization.

The potential of eq.~(\ref{eq:potential_electrostatic}) in therefore found to be relevant to describe the electron-hole bound states (excitons). The use of this potential changes the binding energies of excitons and thus the optical absorption spectrum of the material compared to the use of the Coulomb potential proportional to $r^{-1}$. 

\subsection{Hydrogenic model for excitons through variational principle}

We propose here to describe excitons in 2D materials adopting a hydrogenic model based on the potential of eq.~(\ref{eq:potential_electrostatic}) and on a variational principle with solutions of a 2D hydrogen atom with a potential $\propto r^{-1}$.

The problem to solve for the Hydrogen atom on a 2-brane is the Schrödinger equation:
\begin{equation}
    \bigg( \frac{-\hbar^2}{2m_e} \grad^2 + V(\hat{\vec{r}}) \bigg) \Psi(\hat{\vec{r}} ) = E \Psi(\hat{\vec{r}} )
    \label{eq:schrod}
\end{equation}
with $\grad^2 $ the 2D Laplacian which is given by $\pdv[2]{}{r} + \frac{1}{r} \pdv{}{r} + \frac{1}{r^2} \pdv[2]{}{\theta}$ in polar coordinates, $\Psi(\hat{\vec{r}} )$ the electronic wave-function and $V(\hat{\vec{r}})$ the potential of eq.~(\ref{eq:potential_electrostatic}).

First, when using a potential $\propto r^{-1}$ in eq.~(\ref{eq:schrod}), one can solve the problem exactly, as shown in Refs.~\cite{zaslow_two-dimensional_1967, yang_analytic_1991}. Separating the radial and angular variables, we can write the wave-function as $\Psi(\hat{\vec{r}} ) = R(r) \frac{e^{i l \theta}}{ \sqrt{2\pi}}$, $R(r)$ being the radial wave-function and $l=0, 1, 2, 3, \dots$ the azimuthal quantum number. The radial Schrödinger equation can be written as:
\begin{equation}
    \bigg[ \frac{-\hbar^2}{2m_e} \bigg(  \pdv[2]{}{r} +\frac{1}{r}  \pdv{}{r} -\frac{l^2}{r^2}  \bigg) + V(r) \bigg] R(r) = E R(r)
    \label{eq:radial_schrod}
\end{equation}

A careful treatment of this radial Shrödinger equation for $V(r)\propto r^{-1}$ leads to the following radial wave-functions~\cite{zaslow_two-dimensional_1967, yang_analytic_1991}:
\begin{equation}
    \begin{split}
        R_{n,l}(r)=  \beta f(n,l) e^{\frac{-\beta r}{2}} (\beta r)^{|l|} L_{n+|l|-1}^{2|l|} (\beta r)
    \end{split}
    \label{eq:hydrogen_radial}
\end{equation}
with $f(n,l)=\Bigg[ \frac{(n-1-|l|)!}{((|l|+n-1)!)^3(2n-1)}  \Bigg]^{\frac{1}{2}} $, $n$ a strictly positive integer, $l$ an integer such that $|l|<n-1$, $\beta=\frac{2m_ee^2}{4\pi\epsilon_0  (n-\frac{1}{2})\hbar^2}$ and $L^m_n$ the generalized Laguerre polynomials.

When considering the potential of eq.~(\ref{eq:potential_electrostatic}) in the radial Schrödinger equation~(\ref{eq:hydrogen_radial}), the hydrogen problem is no longer analytically solvable. We therefore used a numerical variational principle (see, e.g., Ref.~\cite{lowe_quantum_2006}) to find the five lowest energy $l=0$ (s-type) states. Our method was implemented in two steps:
\begin{itemize}
    \item wave-functions of the form of eq.~(\ref{eq:hydrogen_radial} were used for the 1s, 2s, \dots, 6s wave-functions and $\beta$ was taken as a variational parameter to numerically minimize the associated energy
    \item a linear variational method was used to minimize the energies by taking linear combinations of the $\beta-$optimized wave-functions of the type eq.~(\ref{eq:hydrogen_radial}).
\end{itemize}

The energy levels of the hydrogen atom on a 2-brane are discussed in Appendix~\ref{sec:app_energies_H}.

When moving from hydrogen atom to a hydrogenic description of excitons, the excitons binding energies are deduced from the hydrogen electron's energies via the following relationship:
\begin{equation}
        E_n^b = -E_n^H \frac{\mu}{m_e}\frac{1}{\epsilon_r^2}
    \label{eq:hydrogenic_energies}
\end{equation}
with $E_n^b$ the excitons' binding energies, $E_n^H$ the hydrogen electron's energies (found from the variational principle), $\mu$ the electron effective mass in the 2D material and $\epsilon_r$ the effective permittivity of the surrounding material (i.e. the bulk in the brane analogy).

The $r_0$ parameter intering the potential also has to be rescaled as follows:
\begin{equation}
        r_0^{exc}=r_0^{H} \epsilon_r \frac{m_e}{\mu}
\end{equation}
with $r_0^{exc}$ the $r_0$ parameter for the exciton's potential and $r_0^{H}$ the $r_0$ parameter for the 2D hydrogen atom. 

\begin{table}[h]
\center
\begin{tabular}{|c|c|c|c|c|}
  \hline
   Orbital & $E_{b,exp}\hspace{0.1cm}[eV]$ & $E_b^K\hspace{0.1cm}[eV]$($r_0=37\buildrel _{\circ} \over {\mathrm{A}}$) &$E_b^{3D}\hspace{0.1cm}[eV]$&$E_b^{2D}\hspace{0.1cm}[eV]$\\
  \hline
  $1s$ & 0.32 & 0.32 &0.089&0.36\\
  $2s$ & 0.16 & 0.14&0.022&0.040 \\
 $3s$ & 0.10 & 0.08 &0.0099&0.014\\
  $4s$ & 0.06 $(\pm0.02)$ & 0.05&0.0056&0.0073 \\
$5s$ & 0.04 $(\pm0.02)$ & 0.03 &0.0036&0.0044\\
  \hline
\end{tabular}
\caption{Comparison of the binding energies for the five first s-type orbitals for each model under consideration. $E_{b,exp}$ are the experimental results from \cite{chernikov_exciton_2014}, $E_b^K$ are the theoretical results from the brane model, while $E_b^{3D}$ and $E_b^{2D}$ are those from usual hydrogenic models in (3+1)-D and (2+1)-D space-times with a classical Coulomb potential.}
\label{tab:results}
\end{table}

\subsection{Application: Excitons of a WS$_{2}$ monolayer}

We now compute excitons' binding energies of a WS$_{2}$ monolayer using the hydrogenic model and illustrate the improvement of the use of the Rytova-Keldysh potential in combination with a variational principle, comparing the theoretical values to the experimental values from Ref.~\cite{chernikov_exciton_2014}. 

The theoretical values were computed using hydrogenic models in pure (3+1)-D or (2+1)-D spacetimes (using a $\propto r^{-1}$ potential) and with our brane model. The effective mass of electrons in WS$_{2}$ is estimated between $0.15 \hspace{0.1cm} m_e$ and $0.22 \hspace{0.1cm} m_e$~\cite{ramasubramaniam_large_2012, shi_quasiparticle_2013, hichri_exciton_2017}. We used a value of $\mu = 0.22 \hspace{0.1cm} m_e$.

For the (3+1)-D and (2+1)-D excitons, a relative permittivity of $\epsilon_r = 5.8$ was chosen~\cite{hichri_exciton_2016}, while for the 2-brane model the bulk permittivity was taken as the mean of the upper and lower permittivities~\cite{mohammadzadeh_excitonic_2013}. In the experiment to which we compare, WS$_{2}$ was deposited on SiO$_{2}$ ($\epsilon_r=2.1$) and surrounded by air ($\epsilon_r=1$), so we used $\epsilon_r=1.55$.

We compare the three hydrogenic models with the experimental data of Ref.~\cite{chernikov_exciton_2014} in Table~\ref{tab:results}. For the brane model, a value of $r_0 = 37  \buildrel _{\circ} \over {\mathrm{A}}$ was chosen to correspond well to the ratios of the experimental energies.

\section{Conclusion}
We adapted the DGPS mechanism, which describes the localization of gauge fields on an infinitely thin brane. In this model, coupled fermions are themselves localized on the brane.  Using this adapted model, we derived an effective potential on the 2-brane. Interestingly, this potential corresponds to the Rytova-Keldysh potential, which is derived through classical electrodynamics and widely used to describe exciton physics in 2D materials.
We demonstrate how the derived potential can provide a novel physical interpretation of exciton binding energies in WS$_{2}$ monolayers. To achieve this, we employ a two-step variational algorithm to calculate the binding energies of various exciton states in 2D materials, using a hydrogen-like model as a basis. The (2+1)-D hydrogen wave functions serve as an ansatz for the variational principle in our approach.
Through a solid-state realization of a braneworld, one can illustrate the relevance of concepts related to the brane approach considered in physics beyond the Standard Model. The derivation of the Rytova-Keldysh potential from a quantum electrodynamics-based one-loop diagram correction suggests a reinterpretation of the $r_0$ parameter. This parameter can be understood as the ratio between (3+1)-D electrodynamics coupling constants and the effective coupling constant on the brane, which emerges from the one-loop correction.
We expect that applying the DGPS mechanism to finite-size or non-static branes could lead to refinements in the effective potential and in the description of excitons in 2D materials that deviate from perfect planar structures.

\begin{appendices}

\section{Green's function in momentum space}
\label{sec:app_green}

We show here the derivation of the expression of the Green's function at eq.~(\ref{Green_FT}), starting from its momentum equation, eq.~(\ref{eq:green_Fourier}).

We first define the free retarded Green's function ($\Tilde{G}_{R,0}(p, y)$) as the solution of eq.~(\ref{Green_FT}) without the term proportional to m:
\begin{equation}
    M_p^2 (p^2 - \partial_y^2) \Tilde{G}_{R,0}(p, y) = \delta (y).
    \label{eq:def_free_green}
\end{equation}

The retarded Green's function can be written from the free retarded Green's function as:
\begin{equation}
    \Tilde{G}_{R}(p, y) = \Tilde{G}_{R,0}(p, y) - \int \dd{y'} \Tilde{G}_{R,0}(p, y-y') Mm p^2 \delta(y') \Tilde{G}_{R}(p, y').
    \label{eq:green_from_free}
\end{equation}

To verify eq.~(\ref{eq:green_from_free}), we apply $M_p^2 (p^2 - \partial_y^2)$ on both sides and get:
\begin{widetext}
\begin{equation}
    \begin{split}
        M_p^2 (p^2 - \partial_y^2) \Tilde{G}_{R}(p, y) & = M_p^2 (p^2 - \partial_y^2) \Tilde{G}_{R}(p, y) -  M_p^2 (p^2 - \partial_y^2) \int \dd{y'} \Tilde{G}_{R,0}(p, y-y') m p^2 \delta(y') \Tilde{G}_{R}(p, y') \\ 
        & = \delta(y) - \int \dd{y'} \delta(y-y') m p^2 \delta(y') \Tilde{G}_{R}(p, y) \\
        & = \delta(y) - m p^2 \delta(y)  \Tilde{G}_{R}(p, y) 
    \end{split}
\end{equation}
\end{widetext}
where the second equality holds from the definition of the free Green's function at eq.~(\ref{eq:def_free_green}) and the third from the integration over y'. The equation last obtained is the same as eq.~(\ref{eq:green_Fourier}), justifying the link between the Green's function and its free part, eq.~(\ref{eq:green_from_free}). Using this link between Green functions and integrating over y', we get:
\begin{equation}
    \Tilde{G}_{R}(p, y)  = \Tilde{G}_{R, 0}(p, y) - \Tilde{G}_{R, 0}(p, y) m p^2 \Tilde{G}_{R}(p, 0)   .
    \label{eq:green_from_free_2}
\end{equation}

Using eq.~(\ref{eq:green_from_free_2}) at $y=0$, we have:
\begin{equation}
    \Tilde{G}_{R}(p, 0) = \frac{\Tilde{G}_{R, 0}(p, 0) }{1+ \Tilde{G}_{R, 0}(p, 0) m p^2}
\end{equation}
and inserting this last equation in eq.~(\ref{eq:green_from_free_2}), we eventually get:
\begin{equation}
\begin{split}
    \Tilde{G}_{R}(p, 0) & = \Tilde{G}_{R,0}(p, 0) \bigg[  1- \frac{m p^2 \Tilde{G}_{R,0}(p, 0)}{ 1 + m p^2 \Tilde{G}_{R,0}(p, 0)}  \bigg] \\
    & =  \Tilde{G}_{R,0}(p, 0) \frac{1}{ 1 + m p^2 \Tilde{G}_{R,0}(p, 0)}
\end{split}
\label{eq:final_green_free_link}
\end{equation}
which is an explicit link between retarded Green's function as a function of the free retarded Green's function. Hence, we are now left with the task to find the free retarded Green's function verifying eq.~(\ref{eq:def_free_green}). To do so, we consider this equation in Fourier space (on the second component):
 \begin{equation}
     M_p^2 (p^2 + q^2) \Tilde{\Tilde{G}}_{R,0} (p,q) = 1
 \end{equation}
 from which $\Tilde{\Tilde{G}}_{R,0} (p,q)$ is immediately deduced. Taking back the inverse Fourier transform, we find:
 \begin{equation}
 \begin{split}
     \Tilde{G}_{R,0} (p,y) & = \int \frac{\dd{q}}{2\pi} e^{iqy} \frac{1}{M_p^2 (p^2 + q^2)} \\
     & = \frac{1}{2\abs{p} M_p^2} e^{-\abs{p} \abs{y}}
 \end{split}
 \end{equation}
 where the last equality can be found in, e.g., Ref.~\cite{gradshteyn_table_2015}. Putting this expression for the free retarded Green's function in eq.~(\ref{eq:final_green_free_link}), we find the expression of the retarded Green's function:
 \begin{equation}
 \Tilde{G}_R(p,y) = \frac{1}{m p^2 + 2 M_p^2 \abs{p}}     e^{-\abs{p} \abs{y}}
 \end{equation}
 which is the eq.~(\ref{Green_FT}) that we wanted to show in this appendix.

\section{Energy levels of an hydrogen atom on a 2-brane}
\label{sec:app_energies_H}

In a (2+1)-D space-time with a $\propto r^{-1}$ potential, the energy levels of the hydrogen atom are found to be proportional to $E_n \propto (n-\frac{1}{2}) ^{-2}$ (see Ref.~\cite{zaslow_two-dimensional_1967}), in contrast its (3+1)-D analog for which $E_n \propto n ^{-2}$ (see Ref.~\cite{cohen-tannoudji_quantum_2020}). The decrease in the energy levels is greater in the (2+1)-D case when $n$ increases. 

\begin{figure}
\includegraphics[width=\columnwidth]{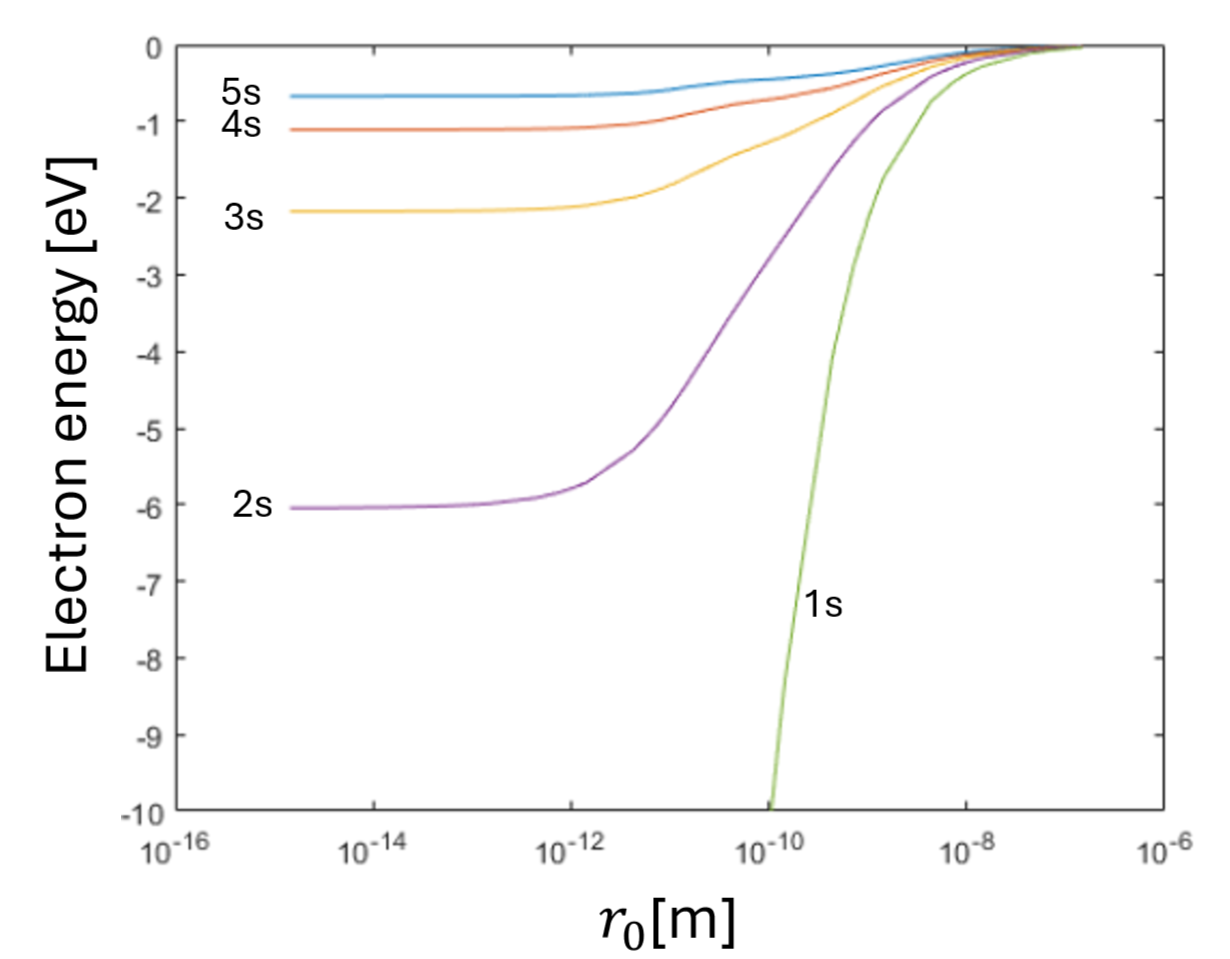}
\caption{Electron energy of an hydrogen atom (1s to 5s) in a 2-brane embedded in a (3+1)-D space-time using the Rytova-Keldysh potential obtained \textit{via} the adapted DGPS mechanism, combined with a variational principle. The energies depends on the $r_0$ parameter which is defined in the main text after eq.~(\ref{eq:potential_electrostatic}) as the ratio between coupling constants in the bulk and on the brane.}
\label{fig:energy_levels_H}
\end{figure}

On a 2-brane, using the potential of eq.~(\ref{eq:potential_electrostatic}) and a numerical variational principle detailed in the main text, the energy levels of 1s to 5s electrons are shown as a function of the $r_0$ parameter (\textit{i.e.} the ratio between the bulk and brane coupling constants) at figure~\ref{fig:energy_levels_H}. At small $r_0$, the $r>>r_0$ limit is achieved, the potential is $\propto r^{-1}$ and the energies tend towards the ones of an hydrogen atom in a (2+1)-D space-time ($\propto (n-\frac{1}{2})^{-2}$). As $r_0$ increases, the ratios between energies decrease and all energies tend to $0$. The $r_0$ parameter in our model is a free parameter that can be adapted to correspond to experiments.

 \end{appendices}

\bibliographystyle{ieeetr}
\bibliography{biblio.bib}


\end{document}